\documentclass{ceurart}

\usepackage[T1]{fontenc}
\usepackage{graphicx}
\usepackage{amsmath,amssymb}
\usepackage{booktabs}
\usepackage{hyperref}
\usepackage{float} % opcional, para usar [H]
\usepackage{caption}

\title{CHARLIE: An On-Premise Multi-Agent Retrieval-Augmented Generation System for Evidential Reasoning in Forensic Science}

\subtitle{Presented at the 1st Workshop on Reasoning with Evidence in Law Enforcement and Forensics (RELAF 2026), co-located with the 21st International Conference on Artificial Intelligence and Law (ICAIL 2026), Singapore, June 2026. This manuscript is the archival version of the work.}

\author[1,2]{Leandro D. Carneiro}
\author[1]{André L. S. Meirelles}
\author[1]{Juliano de A. Gomes}
\author[1,2]{Rafael C. A. Cabral}

\address[1]{Forensic Institute, Civil Police of Federal District, Brazil}
\address[2]{University of Brasília, Brazil}

\conference{RELAF 2026: 1st Workshop on Reasoning with Evidence in Law Enforcement and Forensics, co-located with ICAIL 2026, Singapore, June 2026}
\copyrightyear{2026}
\copyrightclause{Copyright for this paper by its authors. Use permitted under Creative Commons License Attribution 4.0 International (CC BY 4.0).}

\begin{document}

\maketitle

% ------------------------------------------------------
% Abstract
% ------------------------------------------------------
\begin{abstract}
We present Charlie, an on-premise multi-agent Retrieval-Augmented Generation (RAG) system for structured evidential processing in digital forensic environments. Contemporary forensic workflows must handle large volumes of heterogeneous and unstructured documents under strict requirements of traceability, confidentiality, and legal compliance. Charlie addresses this challenge through a controlled agent architecture that combines local retrieval, task decomposition, structured memory, and verification mechanisms. Unlike cloud-based systems, it operates entirely within institutional infrastructure, preserving data sovereignty and evidential integrity. We describe the system’s architecture, including its transition from classical RAG to agent-based orchestration, and demonstrate its application in real-world forensic scenarios. Case studies show that Charlie enables scalable multi-document data extraction and supports longitudinal forensic intelligence generation while maintaining traceability and auditability. Our results indicate that agent-orchestrated, on-premise RAG architectures can effectively support evidential workflows without compromising legal and institutional constraints. Charlie provides a practical and reproducible blueprint for deploying AI systems in high-stakes forensic environments. This manuscript is an archival version of a paper presented at the RELAF 2026 Workshop.
\end{abstract}

% ------------------------------------------------------
% Introduction
% ------------------------------------------------------
\section{Introduction}

Contemporary digital forensics operates under conditions of increasing data saturation. Modern criminal investigations generate large volumes of heterogeneous and predominantly unstructured information, including crime scene reports, laboratory analyses, digital extractions, and geospatial data. These materials are distributed across multiple formats and repositories, often lacking standardized structure and machine-readable organization. As a result, forensic institutions face a growing mismatch between evidential data volume and the bounded cognitive capacity of human analysts.

Forensic analysis requires not only interpretation of individual documents but also structured extraction, cross-document correlation, and identification of recurring entities and patterns, all under strict requirements of traceability, reproducibility, and chain-of-custody compliance. The challenge is therefore to scale evidential processing while preserving procedural integrity.

Recent advances in Large Language Models (LLMs) have demonstrated strong capabilities in natural language understanding, information extraction, and contextual reasoning~\cite{vaswani2017attention, brown2020language, touvron2023llama}. These capabilities suggest potential for supporting large-scale evidential processing. However, naive deployment of LLMs in forensic contexts introduces significant risks. LLMs are probabilistic systems that may produce hallucinated outputs under incomplete context~\cite{ji2023survey}. In addition, context-window limitations constrain multi-document analysis, cloud-based inference raises data sovereignty and confidentiality concerns, and opaque reasoning processes hinder auditability and legal scrutiny.

These constraints are particularly critical in forensic environments, where sensitive data and evidential materials are subject to strict legal and institutional controls. AI systems in such contexts must therefore ensure data sovereignty, traceability, explainability, and human oversight – a blueprint for high-stakes public security institutions worldwide.

To address these requirements, we developed Charlie, an on-premise AI system designed for evidential processing within the Forensic Institute of the Civil Police of the Federal District (Brazil). Charlie is not a general-purpose conversational system but a domain-specific analytical infrastructure supporting structured information extraction, multi-document analysis, and evidential consolidation.

The system operates entirely within institutional infrastructure, ensuring that no data are transmitted externally. It is built on open-source components and open-weight models, promoting transparency and long-term autonomy. Architecturally, Charlie integrates large language models within a multi-agent, retrieval-augmented framework that supports task decomposition, structured memory, and traceable synthesis~\cite{lewis2020rag, yao2022react, zhao2023agents}.

This paper presents the design, architecture, and operational evaluation of Charlie. We describe its evolution from classical RAG to agent-based orchestration, demonstrate its application in real-world forensic scenarios, and analyze its legal and ethical implications. The results indicate that agent-orchestrated, on-premise RAG systems can support scalable evidential processing while preserving institutional control and procedural guarantees.

Charlie contributes to the AI \& Law literature by demonstrating that advanced language models can be integrated into forensic workflows through architectures explicitly aligned with sovereignty, auditability, and human-in-the-loop oversight. 

An earlier version of this work was presented at the RELAF 2026 Workshop, held in conjunction with ICAIL 2026. The present manuscript constitutes the archival version and includes minor revisions for clarity.

% ------------------------------------------------------
% Related Work
% ------------------------------------------------------
\section{Related Work}

Charlie lies at the intersection of five research fields: (i) computational models of evidential reasoning, (ii) probabilistic approaches to legal proof, (iii) knowledge representation in forensic domains, (iv) retrieval-augmented and agent-based architectures for grounded reasoning, and (v) agent architectures.

\subsection{Computational Models of Evidential Reasoning}

AI and Law research have extensively studied formal models of evidential reasoning, particularly argumentation frameworks and hybrid approaches combining logical and probabilistic elements~\cite{prakken2018argumentation}. These models capture the adversarial and defeasible nature of legal reasoning.

However, they typically assume that evidential elements are already structured. In practice, forensic workflows face an upstream bottleneck: processing massive unstructured document volumes. Charlie addresses this preparatory layer by enabling scalable extraction and structuring of evidential data, supporting downstream reasoning models.

\subsection{Probabilistic Legal Modeling}

Bayesian networks have been widely applied to represent evidential dependencies and hypothesis evaluation in legal contexts~\cite{fenton2016risk}. While effective for modeling uncertainty, their construction requires manual identification of relevant variables and relationships.

Charlie complements these approaches by automating large-scale extraction of candidate variables and attributes from forensic documents, facilitating subsequent probabilistic modeling.

\subsection{Knowledge Representation in Forensic Domains}

Knowledge representation efforts, such as CASE~\cite{case2018standard}, aim to formalize investigative data through structured schemas capturing entities, relationships, and provenance.

Charlie aligns with this perspective by producing structured, traceable outputs. Although not an ontology itself, its architecture supports transformation of unstructured narratives into machine-readable representations suitable for interoperability and further analysis.

\subsection{Retrieval-Augmented Generation}

Retrieval-Augmented Generation (RAG) improves factual grounding by conditioning LLM outputs on retrieved documents~\cite{lewis2020rag, gao2023rag}. While widely used in open-domain and enterprise applications, these systems typically operate without constraints related to confidentiality, auditability, or evidential integrity.

Charlie adapts RAG to forensic environments by combining local retrieval, structured memory, and controlled execution, ensuring that all processing remains traceable and institutionally bounded.

\subsection{Agent Architectures}

Recent work on LLM-based agents highlights the role of planning, memory, and tool use in complex task execution~\cite{zhao2023agents, xi2023rise, li2024multiagent}. Frameworks such as ReAct~\cite{yao2022react} enable dynamic interaction with external tools.

However, many agent systems assume open access to external APIs, which is incompatible with forensic constraints. Charlie adopts an agentic approach but restricts actions to controlled internal operations, ensuring reproducibility and compliance with evidential requirements.

\subsection{Positioning}

Charlie contributes by integrating RAG and agent orchestration within a fully on-premise, sovereignty-preserving architecture. It addresses the upstream challenge of transforming unstructured forensic documents into structured, traceable representations, enabling scalable multi-document analysis under evidential constraints.

Rather than replacing formal reasoning models, the system supports their application by bridging document processing and structured evidential representation.

% ------------------------------------------------------
% Design Principles
% ------------------------------------------------------
\section{Design Principles}

Charlie's architecture embeds forensic constraints (data sovereignty, auditability, and procedural integrity) directly into its core design. Charlie was designed under normative, institutional, and procedural constraints intrinsic to forensic practice. Unlike general-purpose AI systems, forensic applications operate within regulated evidential environments where technical architecture must align with legal doctrine, procedural integrity, and institutional governance.

Accordingly, the system is structured around five design principles: (i) data sovereignty, (ii) infrastructural autonomy, (iii) evidential transparency, (iv) structured multi-document reasoning, and (v) modular scalability~\cite{zhao2023agents, xi2023rise}.

\subsection{Data Sovereignty}

Forensic datasets contain sensitive and legally protected information subject to strict regulatory frameworks. Charlie therefore operates entirely on-premise, ensuring that no documents, embeddings, or intermediate outputs are transmitted to external services. All components, including model inference, retrieval, and storage, are executed within institutional infrastructure.

This design preserves confidentiality, reduces exposure to data leakage, and maintains alignment between computational processes and institutional jurisdiction. Data sovereignty is thus treated as an architectural constraint rather than an implementation choice.

\subsection{Open-Source Autonomy}

Charlie is built exclusively on open-source and open-weight technologies, including Qwen3:32B, vLLM~\cite{kwon2023vllm}, LangGraph, and local vector databases. This approach avoids vendor lock-in, enhances transparency, and supports long-term institutional sustainability.

Open infrastructure enables inspection of system behavior and ensures that deployment, maintenance, and evolution remain under institutional control, aligning technological autonomy with governance requirements.

\subsection{Evidential Transparency}

Forensic outputs must be traceable and reproducible. Charlie incorporates transparency at the workflow level by logging query classification, task decomposition, retrieval results, intermediate outputs, and final synthesis steps.

All generated outputs maintain explicit linkage to source documents, preserving the distinction between primary evidence and analytical artifacts. This design supports auditability and aligns with procedural requirements for evidential accountability~\cite{ji2023survey}.

\subsection{Structured Multi-Document Reasoning}

Forensic analysis requires reasoning across document collections rather than isolated texts. Charlie addresses this through task decomposition and structured aggregation, transforming complex instructions into atomic subqueries executed across document sets.

Results are stored in structured memory representations before synthesis, enabling consistent formatting, reducing cross-document contamination, and preserving traceability. This approach circumvents context-window limitations and supports scalable corpus-level analysis~\cite{lewis2020rag}.

\subsection{Scalability and Modularity}

Charlie supports both computational scalability, through parallel execution and optimized inference, and architectural extensibility. The system is designed as a modular framework in which additional components, such as probabilistic reasoning or knowledge representation modules, can be integrated without redesign.

This modularity ensures adaptability to evolving forensic requirements and regulatory environments.

\subsection{Summary}

These principles embed legal and institutional constraints directly into system architecture. Charlie is therefore not a generic LLM application, but a governance-aligned evidential processing system in which technical design is subordinated to procedural legitimacy.

% ------------------------------------------------------
% System Architecture
% ------------------------------------------------------
\section{System Architecture}

Charlie evolved from a classical Retrieval-Augmented Generation (RAG) configuration into a structured multi-agent evidential processing framework~\cite{lewis2020rag, zhao2023agents, xi2023rise}. This transition reflects a shift from single-query response generation to orchestrated evidential workflows.

\subsection{Controlled RAG Foundation}

The system employs a secure, on-premise Retrieval-Augmented Generation (RAG) architecture~\cite{lewis2020rag, gao2023rag}. Documents are preprocessed through text normalization, semantic chunking, and metadata association. Chunks are embedded using the \texttt{nomic-embed-text} model~\cite{nomic2024embed} executed locally via the Ollama framework~\cite{ollama2024}. The resulting embeddings are indexed in a local vector database, enabling efficient similarity-based retrieval while preserving full data sovereignty.

Retrieval follows a two-stage pipeline:

\begin{itemize}
\item Dense retrieval: cosine similarity, top-$k=100$
\item Reranking: \texttt{cross-encoder/ms-marco-MiniLM-L-6-v2}, top-$k=25$
\end{itemize}

Retrieved fragments are incorporated into structured prompts, grounding generation in evidential context and reducing hallucination risk~\cite{ji2023survey}.

However, single-pass RAG exhibited limitations in multi-document extraction, lack of persistent aggregation, and inability to decompose complex tasks.

\subsection{Agent-Orchestrated Processing}

To address the limitations of traditional RAG pipelines in multi-document forensic contexts, Charlie introduces an agent-based orchestration layer~\cite{zhao2023agents, li2024multiagent}, as shown in Figure~\ref{fig:charlie-architecture}. This layer transforms the system from a passive question-answering interface into an active evidential workflow engine capable of structured reasoning over large document corpora.

Incoming queries are first classified into task categories (e.g., direct query, structured extraction, aggregation, or complex analysis), determining the execution strategy. Complex instructions are then decomposed into atomic subqueries, enabling fine-grained processing. For a corpus of $N$ documents and $M$ target attributes, the system generates $N \times M$ subqueries, allowing scalable and systematic extraction across all sources.

Each subquery is executed independently through the RAG pipeline, enabling parallel processing and avoiding the constraints of limited context windows. Retrieved information is stored in structured memory representations, such as tables or key-value mappings, with explicit provenance links to the originating documents. This ensures that all extracted data remains traceable and auditable.

A final synthesis stage consolidates the structured memory into coherent outputs, including tables, JSON objects, or grounded summaries. By separating extraction from synthesis, the architecture improves robustness, reduces error propagation, and enables consistent handling of large-scale evidential datasets.

Overall, this agent-orchestrated architecture integrates classification, decomposition, parallel execution, structured memory, and synthesis into a unified pipeline. This design enables scalable multi-document processing, independence from context-window limitations, consistent structured extraction, and full document-level traceability, making it suitable for operational forensic environments.

\begin{center}
\centering
\includegraphics[width=0.65\linewidth]{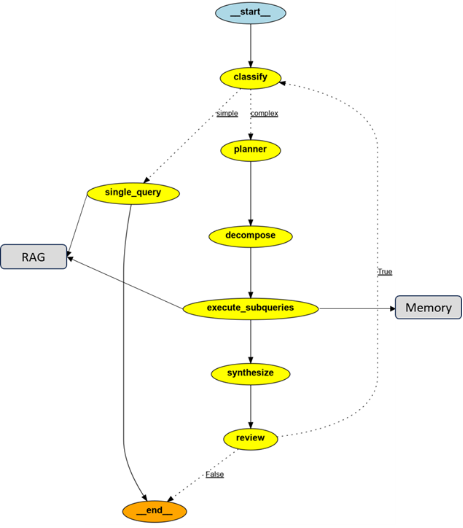}
\captionof{figure}{Charlie agent architecture and evidential workflow graph implemented using LangGraph.}
\label{fig:charlie-architecture}
\end{center}

\subsection{Implementation Details}

Documents are segmented into chunks of approximately 1,000 tokens with an overlap of 100 tokens, balancing contextual coherence with retrieval precision. This configuration was empirically chosen to preserve semantic continuity across segments while minimizing redundancy in the vector index.

Embeddings are generated using the \texttt{nomic-embed-text} model executed locally via the Ollama framework, selected for its multilingual performance and compatibility with on-premise deployment constraints. This choice enables high-quality semantic representations without requiring external API calls, thereby preserving data sovereignty.

Retrieval follows a two-stage pipeline combining dense vector search and cross-encoder reranking. Initially, the system retrieves the top-$k=100$ most relevant chunks based on embedding similarity. These candidates are then reranked using a cross-encoder model (\texttt{ms-marco-MiniLM-L-6-v2}) ~\cite{reimers2019sentencebert} , reducing the set to the top-$k=25$ passages. This approach improves precision by incorporating deeper semantic matching during the reranking stage.

Prompting strategies are tailored to each stage of the agent workflow, with state-specific prompts enforcing structured outputs such as JSON objects, key-value pairs, and explicit null representations. This design ensures consistency across subqueries and facilitates downstream aggregation and validation.

Orchestration is implemented using LangGraph as a directed execution graph, with nodes corresponding to classification, planning, decomposition, execution, memory update, validation, and synthesis. Transitions between nodes are deterministic and auditable, enabling full traceability of the processing pipeline.

Inference is performed using the Qwen3-32B model deployed via vLLM on a cluster of four NVIDIA A40 GPUs with tensor parallelism. This configuration supports large-context processing (up to 131k tokens) and enables high-throughput parallel execution, which is essential for handling multi-document workloads in forensic scenarios.

\subsection{Reproducibility}

The system is fully specified in terms of segmentation, embedding, retrieval, prompting, orchestration, and inference configuration. While datasets are confidential, the architecture is reproducible under equivalent infrastructure.

% ------------------------------------------------------
% Agent Architecture
% ------------------------------------------------------
\section{Agent Architecture}

Charlie implements an agent-based architecture grounded in the perception--reasoning--action paradigm~\cite{zhao2023agents, xi2023rise}. Modern LLM agents extend this model by integrating probabilistic reasoning with retrieval, memory, and tool use~\cite{zhao2023agents, xi2023rise, li2024multiagent}. 

In Charlie, this paradigm is adapted to a controlled evidential environment. The architecture is structured around five components: (i) Brain, (ii) Perception, (iii) Action, (iv) Memory, and (v) Self-Reflection.

\subsection{Brain}

The Brain is an open-weight LLM (Qwen3:32B) deployed locally via vLLM~\cite{kwon2023vllm}. It performs instruction interpretation, subquery generation, contextual reasoning, and output synthesis.

Charlie supports two inference modes: (i) a reasoning mode for complex, multi-step tasks, and (ii) a fast mode for simple grounded queries. This allows dynamic allocation of computational resources.

All reasoning is strictly grounded in retrieved documents. The model does not access external resources, and all inputs are mediated by the Perception layer.

\subsection{Perception}

The Perception layer provides controlled access to evidential data through:

\begin{itemize}
\item Vector-based retrieval
\item Chunk indexing and metadata
\item Structured memory querying
\end{itemize}

For each subquery, relevant document fragments are retrieved from the local corpus~\cite{lewis2020rag, gao2023rag}. This ensures grounding, reduces hallucination risk, and preserves traceability through logged retrieval operations.

Perception also enables access to previously stored structured memory, supporting cross-subquery consistency without requiring full corpus concatenation.

\subsection{Action}

The Action component defines the set of allowed operations. Unlike open-ended agent systems, Charlie restricts actions to predefined internal processes, including subquery generation, retrieval, parallel execution, structured output construction, and memory updates~\cite{yao2022react, zhao2023agents}.

This constraint ensures reproducibility, prevents uncontrolled behavior, and maintains compliance with evidential requirements.

\subsection{Memory}

Charlie implements both short-term working memory and structured persistent memory~\cite{zhao2023agents, li2024multiagent}. Structured memory stores extracted attributes in tabular or key-value formats, maintaining explicit links to source documents.

This enables accumulation of results across documents, avoids redundant processing, and supports grounded synthesis. Memory thus functions as an intermediate layer between raw text and higher-level analysis.

\subsection{Self-Reflection}

A validation layer performs completeness and consistency checks before final synthesis~\cite{zhao2023agents, li2024multiagent}. These include detection of missing fields, structural inconsistencies, ambiguous outputs, and potential hallucination signals~\cite{ji2023survey}.

If issues are identified, the pipeline can be re-executed. This mechanism reduces silent failures and improves robustness in structured extraction tasks.

\subsection{Summary}

By integrating Brain, Perception, Action, Memory, and Self-Reflection, Charlie implements a controlled agent architecture aligned with evidential constraints. The system operates on institutionally bounded data, executes predefined operations, preserves provenance, and validates outputs prior to synthesis.

Rather than acting as an autonomous decision-maker, the agent functions as a structured reasoning component within a governed evidential workflow.

% ------------------------------------------------------
% Operational Evaluation
% ------------------------------------------------------
\section{Operational Evaluation}

Charlie was evaluated through deployment in real-world forensic scenarios within the Forensic Institute. Rather than using open-domain benchmarks, the system was assessed via operational case studies reflecting institutional workflows. The evaluation focused on (i) structured multi-document extraction, (ii) scalability, and (iii) longitudinal intelligence generation.

The objective is not automated adjudication, but architectural support for scalable workflows.

\subsection{Case Study I: Traffic Accident Intelligence}

\textbf{Dataset.} A yearly corpus of approximately 2,000 forensic reports related to traffic accidents, including narrative descriptions, victim data, expert analysis, and geospatial information.

\textbf{Task.} Extraction of structured variables (e.g., incident ID, location, date, collision type, vehicle categories, injury severity) to generate datasets for geospatial and statistical analysis.

\textbf{Processing.} The system decomposed tasks into document-level subqueries, performed retrieval-grounded extraction, and aggregated results into a consolidated dataset~\cite{lewis2020rag, gao2023rag}.

\textbf{Outcome.} The resulting dataset enabled identification of accident clusters, visualization of high-risk segments, and temporal analysis. The system has supported four consecutive years of annual reporting, enabling longitudinal monitoring of traffic patterns.

\subsection{Case Study II: Femicide Pattern Analysis}

\textbf{Dataset.} Forensic reports from femicide investigations in the Federal District, containing heterogeneous narrative descriptions of crime scenes, victim profiles, and offender behavior.

\textbf{Task.} Extraction and aggregation of semantic patterns across five dimensions: instruments used, modus operandi, victim profiles, offender profiles, and substance involvement.

\textbf{Processing.} The agent architecture performed structured subquery-based extraction across documents, followed by corpus-level aggregation~\cite{zhao2023agents, xi2023rise, li2024multiagent}.

\textbf{Outcome.} The system identified recurring patterns, including dominant instrument categories, behavioral dynamics, victim-offender relationships, and contextual factors such as substance use. All outputs maintained traceability to source documents.

\textbf{Impact.} Results supported forensic intelligence generation, investigative prioritization, and evidence-based policy discussions. The system enabled transformation of unstructured reports into structured datasets suitable for population-level analysis.

\subsection{Discussion}

These case studies highlight the distinction between traditional RAG pipelines and agentic architectures. While conventional systems focus on retrieval and summarization, Charlie enables structured multi-document reasoning with aggregation and traceability.

Key properties observed include:

\begin{itemize}
\item Scalability through parallel subquery execution
\item Grounding via retrieval-based generation
\item Traceability through structured memory and provenance linkage
\item Robustness via completeness verification
\item Operational compatibility with institutional workflows
\end{itemize}

Although qualitative, these results demonstrate that agent-orchestrated, on-premise RAG architectures can support evidential processing at institutional scale. Future work will incorporate quantitative evaluation metrics and benchmarking protocols.

% -----------------------------------------------------
% Legal and Ehtical Considerations
% ------------------------------------------------------
\section{Legal and Ethical Considerations}

The deployment of artificial intelligence systems in forensic and law enforcement contexts raises distinct legal and ethical challenges~\cite{ji2023survey}. Unlike many commercial or enterprise AI applications, forensic systems operate within judicial ecosystems where outputs may influence investigative decisions, prosecutorial strategies, or courtroom proceedings. Consequently, technical design choices must be evaluated not only for performance but also for compliance with procedural guarantees, evidential integrity standards, and fundamental rights protections.

Charlie was developed under the premise that AI systems in evidential contexts must remain subordinate to legal doctrine and human expertise. This section outlines the primary legal and ethical considerations informing its architecture.

\subsection{Preservation of Chain-of-Custody and Evidential Integrity}

Chain-of-custody principles require that evidential materials remain traceable from acquisition through analysis and presentation. Any transformation or processing step must be reconstructable and attributable.

Charlie does not modify original forensic documents. Instead, it operates as an analytical layer that produces derived outputs (structured tables, summaries, or extracted attributes) while preserving explicit references to source documents. All retrieval steps, subqueries, and intermediate outputs are logged, enabling reconstruction of how a given structured result was obtained.

Importantly, Charlie’s outputs are not themselves evidentiary artifacts but analytical aids. The authoritative evidential material remains the original document, which can always be consulted for verification. This distinction preserves procedural integrity and prevents conflation between AI-generated synthesis and primary evidence.

\subsection{Data Protection and Confidentiality}

Forensic datasets frequently contain sensitive personal data, including identifying information, medical details, and protected investigative content. Data protection regimes in many jurisdictions impose strict constraints on processing, transfer, and storage of such data.

Charlie addresses these concerns through complete on-premise execution. No data are transmitted to external cloud services, and no third-party APIs are invoked during inference. By confining processing to secure institutional infrastructure, the system reduces risks associated with cross-border data flows, external vendor access, or unauthorized retention.

Furthermore, embedding generation and vector indexing are performed locally, ensuring that even intermediate semantic representations remain under institutional control. From a data governance perspective, the architecture maintains alignment between technical processing boundaries and institutional jurisdictional authority.

\subsection{Human-in-the-Loop Oversight}

A central ethical principle in forensic AI deployment is that automated systems must not replace expert judgment~\cite{zhao2023agents, xi2023rise}. Forensic analysis involves interpretative reasoning, contextual evaluation, and normative assessment that cannot be delegated to probabilistic models.

Charlie was explicitly designed as a decision-support tool rather than an autonomous decision-maker. The system:

\begin{itemize}
\item Performs structured extraction and consolidation
\item Does not issue conclusions regarding guilt, liability, or legal responsibility
\item Does not automatically generate binding determinations
\end{itemize}

Human analysts review all outputs before incorporation into official workflows. In this sense, Charlie operationalizes a human-in-the-loop paradigm in which AI augments cognitive capacity without supplanting professional accountability.

\subsection{Transparency and Explainability}

Explainability in LLM systems is a complex and unresolved research challenge~\cite{ji2023survey}. However, procedural transparency can be strengthened even when internal neural representations remain opaque.

Charlie enhances explainability at the workflow level by:

\begin{itemize}
\item Logging retrieval decisions
\item Recording subquery generation
\item Preserving structured intermediate outputs
\item Maintaining explicit linkage between extracted values and source fragments
\end{itemize}

This approach does not claim full interpretability of the model’s internal token-level reasoning. Instead, it provides practical traceability sufficient for forensic audit and judicial scrutiny. Analysts can verify which textual fragments supported each extracted attribute and assess whether contextual nuance was appropriately captured.

Transparency is therefore achieved through architectural design rather than model introspection.

\subsection{Risk Mitigation: Hallucination and Over-Reliance}

LLMs are known to produce plausible but fabricated outputs under certain conditions~\cite{ji2023survey}. In evidential contexts, hallucinated information poses serious risks, including contamination of investigative narratives or distortion of structured datasets.

Charlie mitigates hallucination risk through multiple mechanisms:

\begin{enumerate}
\item Retrieval-grounded generation limits outputs to semantically retrieved fragments~\cite{lewis2020rag}
\item Structured decomposition reduces ambiguity in complex tasks~\cite{zhao2023agents}
\item Explicit null reporting prevents silent fabrication of missing values
\item Human verification ensures final oversight
\end{enumerate}

Nevertheless, the system cannot eliminate probabilistic error entirely. Ethical deployment therefore requires ongoing evaluation, conservative integration into workflows, and continuous user training to prevent over-reliance on automated outputs.

\subsection{Normative Positioning}

From a normative standpoint, Charlie embodies a governance-aligned approach to forensic AI. It rejects the model of opaque, cloud-hosted AI systems operating beyond institutional oversight. Instead, it integrates advanced generative capabilities within a bounded, auditable, and sovereignty-preserving architecture.

The system’s legal and ethical posture can be summarized as follows:

\begin{itemize}
\item AI augments, but does not replace, expert reasoning
\item Analytical outputs remain subordinate to primary evidence
\item All processing remains within institutional jurisdiction
\item Procedural traceability is embedded into technical architecture
\end{itemize}

By aligning technological design with evidential doctrine, Charlie seeks to demonstrate that generative AI can be responsibly integrated into forensic practice without compromising legality, transparency, or human accountability.

% ------------------------------------------------------
% Limitations
% ------------------------------------------------------
\section{Limitations}

Despite its operational effectiveness, Charlie presents technical, methodological, and normative limitations that must be acknowledged for responsible deployment. The system relies heavily on semantic retrieval to ground generation, which reduces hallucination risk but introduces sensitivity to retrieval quality~\cite{lewis2020rag, gao2023rag}. Missing or poorly ranked fragments may lead to incomplete or null extractions, even when relevant information exists in the corpus.

The use of chunk-based segmentation further introduces challenges related to retrieval noise and context fragmentation. Semantically dependent information may be distributed across multiple segments, while similarity-based retrieval can surface contextually irrelevant passages. Balancing chunk size, overlap, and retrieval depth ($k$) remains an open optimization problem~\cite{gao2023rag}.

At the modeling level, the reasoning core is a probabilistic language model rather than a formally verified reasoning system~\cite{ji2023survey}. Consequently, linguistic ambiguity, domain-specific terminology, and implicit context may not always be correctly interpreted. While self-reflection mechanisms improve structural consistency, they do not eliminate epistemic uncertainty.

From an infrastructure perspective, the fully on-premise deployment model requires substantial computational resources, particularly for large-scale models such as Qwen3:32B~\cite{kwon2023vllm}. Although this design ensures data sovereignty and compliance with evidential constraints, it shifts the burden of infrastructure provisioning, maintenance, and optimization to the institution.

In addition, Charlie currently focuses on structured extraction and does not yet incorporate formal evidential reasoning mechanisms such as Bayesian inference or argumentation frameworks. As such, it supports evidential preparation rather than formal proof modeling~\cite{zhao2023agents, li2024multiagent}. Finally, the present evaluation is based on operational case studies rather than controlled benchmarks, which reflects real-world applicability but limits quantitative comparison across methods.

Overall, Charlie remains dependent on retrieval quality, constrained by probabilistic modeling, computationally demanding, and primarily oriented toward evidential preparation. These limitations define the boundaries of its application while highlighting directions for future improvement.

% ------------------------------------------------------
% Future Work
% ------------------------------------------------------
\section{Future Work}

Charlie was designed as a modular research infrastructure for advancing evidential reasoning systems. While its current implementation focuses on structured extraction and consolidation, future work aims to extend its capabilities toward formal reasoning, improved retrieval, enhanced interpretability, and broader operational deployment.

A central direction involves the integration of formal evidential reasoning mechanisms operating on structured outputs. Bayesian models could leverage extracted attributes as variables to support probabilistic inference, hypothesis evaluation, and sensitivity analysis~\cite{fenton2016risk}. In parallel, argumentation frameworks may enable the automatic construction of argument graphs, identification of evidential conflicts, and structured comparison of competing narratives~\cite{prakken2018argumentation}. Together, these approaches would elevate the system from structured information extraction to formal decision-support.

Although Charlie already provides workflow-level transparency through its multi-agent architecture, further improvements in explainability are required for operational adoption. Future interfaces will focus on retrieval trace visualization, explicit source highlighting, confidence indicators, and completeness diagnostics, improving both usability and trustworthiness~\cite{ji2023survey}.

Finally, as regulatory frameworks for artificial intelligence continue to evolve, Charlie will incorporate governance-aligned features such as automated documentation, risk assessment modules, and model auditing capabilities, ensuring compliance with emerging legal and ethical requirements. Overall, future development will focus on integrating formal reasoning, improving retrieval robustness, enhancing explainability, and establishing standardized evaluation methodologies. Charlie thus serves not only as an operational system but also as a research platform for advancing responsible AI in forensic contexts~\cite{zhao2023agents, li2024multiagent}.

% ------------------------------------------------------
% Conclusion
% ------------------------------------------------------
\section{Conclusion}

This paper introduced Charlie, an on-premise, multi-agent Retrieval-Augmented Generation system designed to support structured evidential processing within forensic institutions~\cite{lewis2020rag, zhao2023agents}. Developed and deployed at Brazil's Forensic Institute, Charlie addresses a fundamental operational challenge: the exponential growth of heterogeneous forensic documentation under stringent procedural and legal constraints.

Unlike generic conversational AI systems, Charlie was architected under explicit normative and institutional design principles. Its fully local deployment ensures data sovereignty and confidentiality preservation. Its open-source stack enhances infrastructural autonomy and long-term sustainability. Its structured agent architecture integrates controlled retrieval, task decomposition, parallel execution, persistent memory, and self-reflection mechanisms, transforming large language models into components of a traceable evidential workflow~\cite{zhao2023agents, li2024multiagent}.

Operational case studies demonstrated the system’s capacity to perform structured multi-document extraction and longitudinal forensic intelligence consolidation across large corpora. These results illustrate that agent-orchestrated RAG architectures can scale beyond isolated question answering to support systematic evidential consolidation~\cite{lewis2020rag, gao2023rag}. Importantly, Charlie does not replace expert reasoning. Instead, it augments forensic workflows while preserving human-in-the-loop oversight, chain-of-custody integrity, and document-level traceability.

Currently, there are no widely available solutions capable of processing thousands of forensic documents to automatically construct structured evidential datasets under strict institutional and legal constraints. In this sense, Charlie fills a critical technological and operational gap.

From a broader AI \& Law perspective, Charlie contributes to ongoing discussions about how generative AI can be responsibly integrated into judicial and law enforcement contexts~\cite{ji2023survey}. Rather than pursuing unconstrained automation, the system embodies a governance-aligned approach in which technical architecture is subordinated to legal doctrine and institutional sovereignty. By embedding auditability, constrained action, and structured reasoning into its design, Charlie demonstrates that advanced language models can be deployed in high-stakes evidential environments without sacrificing procedural legitimacy.

At the same time, the system’s current scope remains bounded to evidential preparation and structured extraction. Future integration with formal probabilistic reasoning frameworks, argumentation models, and knowledge representation standards will further align the architecture with computational models of legal proof~\cite{fenton2016risk, prakken2018argumentation}.

Charlie thus represents both an operational system and a research platform. It provides a replicable blueprint for forensic institutions seeking to adopt AI technologies responsibly, demonstrating that sovereignty-preserving, transparent, and scalable multi-agent architectures are viable within law enforcement environments. In doing so, it advances the interdisciplinary dialogue on how artificial intelligence can support evidential reasoning while respecting the legal and ethical foundations of forensic practice.

\section{Availability}
This manuscript is available as an arXiv preprint to provide a permanent, citable archival version of the work presented at RELAF 2026.

\section{Acknowledgements}
The authors thank the organizers of the RELAF 2026 Workshop for providing an interdisciplinary forum for discussing AI applications in law enforcement and forensic science.

\bibliography{charlie_relaf_arxiv}

\end{document}